\RequirePackage{ifpdf}
\ifpdf 
\documentclass[pdftex]{sigma}
\else
\documentclass{sigma}
\fi

\begin{document}

\renewcommand{\thefootnote}{$\star$}

\renewcommand{\PaperNumber}{009}

\FirstPageHeading

\ShortArticleName{${\cal P}{\cal T}$ Symmetric Schr\"odinger Operators: Reality of the Perturbed Eigenvalues}

\ArticleName{$\boldsymbol{{\cal P}{\cal T}}$ Symmetric Schr\"odinger Operators:\\ Reality of the Perturbed Eigenvalues\footnote{This paper is a
contribution to the Proceedings of the 5-th Microconference
``Analytic and Algebraic Me\-thods~V''. The full collection is
available at
\href{http://www.emis.de/journals/SIGMA/Prague2009.html}{http://www.emis.de/journals/SIGMA/Prague2009.html}}}

\Author{Emanuela CALICETI~$^\dag$, Francesco CANNATA~$^\ddag$ and Sandro GRAFFI~$^\dag$}

\AuthorNameForHeading{E. Caliceti, F. Cannata and S. Graf\/f\/i}

\Address{$^\dag$~Dipartimento di Matematica, Universit\`a di Bologna, and INFN, Bologna, Italy}
\EmailD{\href{mailto:caliceti@dm.unibo.it}{caliceti@dm.unibo.it}, \href{mailto:graffi@dm.unibo.it}{graffi@dm.unibo.it}}

\Address{$^\ddag$~INFN, Via Irnerio 46, 40126 Bologna, Italy}
\EmailD{\href{mailto:Francesco.Cannata@bo.infn.it}{Francesco.Cannata@bo.infn.it}}

\ArticleDates{Received November 03, 2009, in f\/inal form January 14, 2010;  Published online January 20, 2010}

\Abstract{We prove the reality of the perturbed eigenvalues of some ${\cal P}{\cal T}$ symmetric Hamiltonians of physical interest by means of stability methods. In particular we study 2-dimensional generalized harmonic oscillators with polynomial perturbation and the one-dimensional $x^2(ix)^{\epsilon}$ for $-1<\epsilon<0$.}


\Classification{47A55; 47A75; 81Q15; 34L40; 35J10}

\section{Introduction and statement of the results}\label{section1}

A basic mathematical question in the framework of ${\cal P}{\cal T}$ symmetric quantum mechanics concerns the reality of the spectrum of the considered Hamiltonian. One crucial issue is to formulate conditions for the reality of (part of) the spectrum of Hamiltonians depending on a real parameter~$\epsilon$ of the form
\begin{equation}
\label{Ham}
H(\epsilon) = H_0 + \epsilon V , \qquad \epsilon\in{\mathbb R}.
\end{equation}
In \cite{CGS1, CGS2} results have been obtained in the case when the unperturbed Hamiltonian $H_0$ is self\-adjoint with discrete spectrum, and in \cite{CCG1} in the more general case when $H_0$ is ${\cal P}{\cal T}$ symmetric but not necessarily selfadjoint, with discrete spectrum. The common framework of these papers is perturbation theory, which allows to obtain the reality result only for small values of the coupling constant $\epsilon$. However this is not a limitation: in many cases as $\epsilon$ varies, critical phenomena occur, such as a spontaneous breaking of the ${\cal P}{\cal T}$ symmetry and the appearance of complex eigenvalues caused by the crossing of energy levels of $H_0$. The most elementary example of this phenomenon is represented by a $2\times 2$ matrix of the form
\begin{equation}
\label{matrix}
H_1(\epsilon) = \left(\begin{array}{cc} e_1 & i\epsilon
\\
i\epsilon & e_2
\end{array}\right)
\end{equation}
with $e_1,  e_2\in{\mathbb R}$, recently examined in \cite{Rotter, Guo}. $H_1(\epsilon)$ is ${\cal P}{\cal T}$ symmetric, i.e.\ $H_1(\epsilon)({\cal P}{\cal T}) = ({\cal P}{\cal T})H_1(\epsilon)$, if the parity operator ${\cal P}$ is the unitary involution def\/ined by
\[
{\cal P} = \left(\begin{array}{cc} 1 & 0
\\
0 & -1
\end{array}\right)
\]
and ${\cal T}$ is the antilinear involution of complex conjugation. The eigenvalues of $H_1(\epsilon)$ are
\[
\lambda(\epsilon) = \tfrac{1}{2}(e_1+e_2) \pm \tfrac{1}{2}\sqrt {(e_1-e_2)^2 - 4\epsilon^2}
\]
and they are real if and only if $|\epsilon|\leq\frac{|e_1-e_2|}{2}$. The matrix (\ref{matrix}) can be rewritten in the perturbative form~(\ref{Ham})
\[
H_1(\epsilon) = \left(\begin{array}{cc} e_1 & 0
\\
0 & e_2
\end{array}\right) + i\epsilon\left(\begin{array}{cc} 0 & 1
\\
1 & 0
\end{array}\right),
\]
so that it can be put in the framework of Theorem 1.2 of \cite{CGS1}, whose applicability goes far beyond the case of f\/inite dimensional operators (it includes in fact classes of Schr\"odinger operators with bounded perturbation potentials, also in dimension greater than one): this theorem ensures the reality of the spectrum of $H_1(\epsilon)$  for $|\epsilon|<\frac{|e_1-e_2|}{2\|W\|}$ where $\|W\|$ denotes the norm of the bounded perturbation operator $W$, which in the present case is $1$ since $W=\left(\begin{array}{cc} 0 & 1
\\
1 & 0
\end{array}\right)$. The conclusion is that the perturbation argument of \cite{CGS1} yields actually the reality of the spectrum for all the allowed values of $\epsilon$.

In a similar way one can analyze ${\cal P}{\cal T}$ symmetric matrices of the form
\begin{equation}
\label{matrix2}
H_1'(\epsilon) = \left(\begin{array}{cc} e+i\epsilon & b
\\
b & e-i\epsilon
\end{array}\right) = \left(\begin{array}{cc} e & b
\\
b & e
\end{array}\right) + i\epsilon\left(\begin{array}{cc} 1 & 0
\\
0 & -1
\end{array}\right)
\end{equation}
with $e, b\in{\mathbb R}$, which are ${\cal P}{\cal T}$ symmetric with
\[
{\cal P} = \left(\begin{array}{cc} 0 & 1
\\
1 & 0
\end{array}\right).
\]
For a physical interpretation of these matrices see~\cite{Rotter, Guo} for applications to optics, and Section~\ref{section4} of the present paper for an application to classical mechanics. Complex Hamiltonians of type~(\ref{matrix}) are a particular case of those considered in~\cite{Seyranian1}. In turn the Hamiltonians (\ref{matrix2}) are a particular case of those examined in~\cite{Seyranian2}.

The results obtained in \cite{CGS1, CGS2} deal with the case when the perturbation potential $V$ is bounded and, under suitable assumptions on $H_0$ and $V$, they guarantee the reality of the entire spectrum of $H(\epsilon)$ for $|\epsilon|<\delta/\|W\|$, where $\delta=\frac{1}{2}\inf\limits_{n\neq m}|\lambda_n-\lambda_m|$ and $\sigma(H_0)=\{\lambda_n:n\in{\mathbb N}\}$ is the (discrete) spectrum of $H_0$. In some cases, not only the reality of the eigenvalues can be proved, but also the similarity to a selfadjoint operator \cite{Albeverio1, Albeverio2}. In \cite{CCG1} the authors analyze the case when~$V$ is unbounded (and in general not even relatively bounded with respect to~$H_0$), obtaining a~weaker result which guarantees that the perturbed eigenvalues of $H(\epsilon)$, converging to those of~$H_0$ as $\epsilon\to 0$, are real for $|\epsilon|$ suf\/f\/iciently small. In \cite{CCG2} this result has been extended to classes of Hamiltonians of the form
\[
H(\epsilon) = H + iW_{\epsilon}, \qquad \epsilon\geq 0
\]
acting in $L^2({\mathbb R})$, where $H=-\frac{d^2}{dx^2}+x^2$ denotes the operator associated with the one dimensional harmonic oscillator and $W_{\epsilon}\in C^0({\mathbb R})$ is an odd real-valued function: $W_{\epsilon}(-x)=-W_{\epsilon}(x)\in{\mathbb R}$, $\forall \, x\in{\mathbb R}$, $\forall\, \epsilon\geq 0$. Such generalization of the result of \cite{CCG1} is based on the fact that in $W_{\epsilon}$ the dependence on the perturbation parameter $\epsilon$ is the most general one and not only of linear type as in~(\ref{Ham}).
The aim of this paper is to show how simple extensions of the results of  \cite{CCG1, CCG2} allow us to treat models of considerable interest in the context of ${\cal P}{\cal T}$ symmetric quantum mechanics. More precisely, extending the results of~\cite{CCG1} to the case of Schr\"odinger  operators in dimension greater than $1$ it is possible to examine models of the following type
\begin{equation}\label{oscillatori}
H_2(\epsilon) = -\frac{d^2}{dx_1^2} - \frac{d^2}{dx_2^2} + \omega_1^2x_1^2 + \omega_2^2x_2^2 + i\epsilon x_1^rx_2^s , \qquad x=(x_1,x_2)\in{\mathbb R}^2 ,
\end{equation}
where we assume $r,s\in{\mathbb N}$, $r+s$ odd, and $\omega_1, \omega_2>0$. This type of Hamiltonians, which represent a natural generalization of the quantum Henon--Heiles model, has been studied also in \cite{Nanayakkara, Czech}. We will analyze the Hamiltonian (\ref{oscillatori}) in Section~\ref{section2} and we will prove the reality of the eigenvalues $\lambda_{n_1, n_2}(\epsilon)$ of $H_2(\epsilon)$ generated by the unperturbed ones, i.e.\ by the eigenvalues $\lambda_{n_1, n_2}(0) = (2n_1+1)\omega_1+(2n_2+1)\omega_2$, $\forall\, n_1, n_2\in{\mathbb N}_0:=\{0, 1, 2,\dots\}$, such that $\lambda_{n_1, n_2}(\epsilon)\to\lambda_{n_1, n_2}(0)$ as $\epsilon\to 0$, in the case of non-resonant frequencies $\omega_1$, $\omega_2$.

Then, generalizing also the results of \cite{CCG2} we will examine the basic model
\begin{equation}
\label{Tateo}
H_3(\epsilon) = -\frac{d^2}{dx^2} + x^2(ix)^{\epsilon} , \qquad -1<\epsilon<1.
\end{equation}
The Hamiltonian (\ref{Tateo}) has been examined and its physical interest discussed by several authors \cite{BeBo, Tateo2, Tateo3, BeJo} and the reality of its spectrum for $\epsilon>0$ has been proved in \cite{DDT} (see also \cite{Shin}); for $\epsilon<0$ numerical results indicate the appearance of complex eigenvalues (for a recent discussion see e.g.~\cite{Tateo2}), in the sense that, as $|\epsilon|$ increases while $\epsilon$ moves from $0$ to $-1$, the perturbed eigenvalues of $H_3(\epsilon)$, generated by the unperturbed ones $\lambda_n(0)=2n+1$, $\forall\, n\in{\mathbb N}_0$, become complex. In Section~\ref{section3} we will prove that as long as $|\epsilon|$ stays suf\/f\/iciently small, those eigenvalues remain real. We will actually treat the case of small $|\epsilon|$ independently of the sign of $\epsilon$: for $\epsilon>0$ we recover the known result proved in \cite{DDT}.

\begin{remark}
\label{R1}
As in \cite{CCG1, CCG2}, the results of Sections~\ref{section2} and~\ref{section3} are obtained by proving the stability of the unperturbed eigenvalues. For the notion of stability see e.g.~\cite{Kato, RS}. In particular, let us recall that if $\{H(\epsilon): |\epsilon|<\epsilon_0\}$ is a family of closed ${\cal P}{\cal T}$ symmetric operators in a Hilbert space~${\cal H}$ and if a simple (i.e., non degenerate) eigenvalue $\lambda$ of $H(0)$ is stable with respect to the family $\{H(\epsilon): |\epsilon|<\epsilon_0\}$, then for $|\epsilon|$ suf\/f\/iciently small there exists one and only one eigenvalue $\lambda(\epsilon)$ of~$H(\epsilon)$ near~$\lambda$ such that
\[
\lambda(\epsilon)\to\lambda,\qquad {\rm as}\ \  \epsilon\to 0.
\]
Now, recalling that the eigenvalues of a ${\cal P}{\cal T}$ symmetric operator come in pairs of complex conjugate values, we can assert that the uniqueness of $\lambda(\epsilon)$ implies its reality: in fact, if $\lambda(\epsilon)$ is not real then there are two distinct eigenvalues, $\lambda(\epsilon)$ and $\overline{\lambda(\epsilon)}$, and not just one as stated above.
\end{remark}

Assume now that ${\cal H}$$=L^2({\mathbb R}^d)$, $d\geq1$, and that $H(\epsilon)$, $|\epsilon|<\epsilon_0$, is formally given by
\[
H(\epsilon) = p^2 + V_1 + W_{\epsilon},
\]
where $p^2=-\Delta=-\sum\limits_{k=1}^d\frac{d^2}{dx_k^2}$, $x=(x_1,\dots,x_k)\in{\mathbb R}^d$ and $V_1,  W_{\epsilon}\in L^2_{\rm loc}({\mathbb R}^d)$. Moreover we assume that $H(\epsilon)$ has discrete spectrum, i.e.\ each point in the spectrum is an isolated eigenvalue with f\/inite multiplicity (for a review of the notion of multiplicity of an eigenvalue see also~\cite{CCG1}). Let $H^*(\epsilon)$ denote the adjoint of $H(\epsilon)$ and assume that both $H(\epsilon)$ and $H^*(\epsilon)$ have $C^{\infty} _0({\mathbb R}^d)$ as a core.

As recalled in \cite{CCG2} the stability of the eigenvalues of $H(0)$ in the framework of the stability theory developed by Hunziker and Vock in \cite{HV} holds if the following three conditions are satisf\/ied.
\begin{enumerate}\it \itemsep=0pt
\item[$(1)$]
For all $u\in C_0^{\infty}({\mathbb R}^d)$
\begin{equation}\label{continuityA}
\lim_{\epsilon\to 0} \|H(\epsilon)u - H(0)u\| = 0
\end{equation}
and
\begin{equation}\label{continuityB}
\lim_{\epsilon\to 0} \|H^*(\epsilon)u - H^*(0)u\| = 0.
\end{equation}
\item[$(2)$]
There exist constants $a, b>0$ and $\gamma$, $|\gamma|<\pi/2$, such that
\begin{equation}\label{gamma}
\langle u, p^2u\rangle \leq a\{\cos \gamma \Re\langle u, H(\epsilon)u\rangle + \sin \gamma \Im\langle u, H(\epsilon)u\rangle + b\langle u, u\rangle\}
\end{equation}
for all $u\in C_0^{\infty}({\mathbb R}^d)$.
\item[$(3)$]
For any $z\in {\mathbb C}$, there exist positive constants $\delta$, $n_0$ and $\epsilon_1$ such that
\begin{equation}\label{distance}
d_n(z, H(\epsilon)) \geq \delta>0,
\end{equation}
for all $n>n_0$ and $|\epsilon|<\epsilon_1$, where $d_n(z, H(\epsilon)):= {\rm dist}(z, {\cal N}_n(\epsilon))$ and
\[
{\cal N}_n(\epsilon):= \{\langle u, H(\epsilon)u\rangle : u\in D(H(\epsilon)), \|u\|=1, u(x)=0\; {\rm for}\; |x|<n\}
\]
is the so called ``numerical range at infinity''  $($see {\rm \cite{CCG1})}.
\end{enumerate}
Here $D(H(\epsilon))$ denotes the domain of $H(\epsilon)$ and $\langle u,v\rangle:=\int_{{\mathbb R}^d}u(x)\overline{v(x)}dx$ denotes the standard scalar product in $L^2({\mathbb R}^d)$. For future reference we summarize this result in the following criterion, proved in \cite{HV} (see also \cite{CCG2}).

\begin{proposition}
\label{criterion}
The eigenvalues of $H(0)$ are stable w.r.t.\ the family $H(\epsilon)$, $|\epsilon|<\epsilon_0$, if the above conditions $(1)$--$(3)$ are satisfied.
\end{proposition}

In Sections~\ref{section2} and~\ref{section3} we will prove that the operator families $H_2(\epsilon)$ and $H_3(\epsilon)$ satisfy this criterion and, in view of Remark \ref{R1} above, this is enough to ensure the reality of the perturbed eigenvalues of $H_2(\epsilon)$ and $H_3(\epsilon)$.

\section[The operator family $H_2(\epsilon)$: stability and reality of the perturbed eigenvalues]{The operator family $\boldsymbol{H_2(\epsilon)}$: stability and reality\\ of the perturbed eigenvalues}
\label{section2}

The Schr\"odinger operator $H_2(\epsilon)$, $\epsilon\in{\mathbb R}$, introduced in Section~\ref{section1} is def\/ined as the closure in~$L^2({\mathbb R}^2)$ of the minimal operator def\/ined on $C_0^{\infty}({\mathbb R}^2)$, we denote its domain $D(H_2(\epsilon))$, and its adjoint is given by $H_2(-\epsilon)$. It has discrete spectrum (see e.g.~\cite{RS}, Theorem XIII.64; we note that the extension to the present non-selfadjoint case is straightforward) and the unperturbed eigenvalues, i.e.\ the eigenvalues of $H_2(0)$, are given by
\[
\lambda_{n_1, n_2}(0) = (2n_1+1)\omega_1+(2n_2+1)\omega_2,\qquad\forall \, n_1, n_2\in{\mathbb N}_0.
\]
We assume that the frequencies $\omega_1, \omega_2>0$ are non-resonant, i.e.\ the equation $k_1\omega_1+k_2\omega_2=0$, $k_1, k_2\in{\mathbb Z}$, is satisf\/ied if and only if $k_1, k_2=0$. Then each eigenvalue $\lambda_{n_1, n_2}(0)$ is simple, i.e.\ the corresponding eigenspace has dimension~1. Moreover we assume that $r, s\in{\mathbb N}$ are not both even. Then $H_2(\epsilon)$ is ${\cal P}{\cal T}$ symmetric, where ${\cal T}$ is complex conjugation and the parity operator~${\cal P}$ is def\/ined as follows: if~$r$ and~$s$ are both odd~${\cal P}$ changes the sign of one coordinate, i.e.\ it can be either
$
({\cal P}_1u)(x_1,x_2) = u(-x_1,x_2)$ or $({\cal P}_2u)(x_1,x_2) = u(x_1,-x_2)$, $\forall\, x=(x_1,x_2)\in{\mathbb R}^2$, $\forall\, u\in L^2({\mathbb R}^2)$. If $r$ and $s$ are not both odd a suitable choice for ${\cal P}$ is $({\cal P}_3u)(x) = u(-x)$, $\forall \,x\in{\mathbb R}^2$, $\forall\, u\in L^2({\mathbb R}^2)$.

We now prove that conditions (1)--(3) are satisf\/ied.
\begin{proposition}
\label{prop2}
Conditions $(1)$--$(3)$ are satisfied by the operator family $H_2(\epsilon)$, $\epsilon\in{\mathbb R}$.
\end{proposition}

\begin{proof}
(1) $\forall\, u\in C_0^{\infty}({\mathbb R}^2)$ let $K={\rm supp}(u)$ denote the support of $u$; since $K$ is compact in ${\mathbb R}^2$, there exists a constant $M>0$ such that $|x_1^rx_2^s|\leq M$, $\forall\, x=(x_1,x_2)\in K$. Therefore{\samepage
\begin{gather*}
\|H_2(\epsilon)u - H_2(0)u\|^2 = \|H_2^*(\epsilon)u - H_2^*(0)u\|^2 = \int_K|i\epsilon x_1^rx_2^su(x)|^2dx
\nonumber\\
\phantom{\|H_2(\epsilon)u - H_2(0)u\|^2}{} \leq \epsilon^2M^2\int_K|u(x)|^2dx = \epsilon^2M^2\|u\|^2,\qquad \forall\epsilon\in {\mathbb R},\nonumber
\end{gather*}
and this yields (\ref{continuityA}) and (\ref{continuityB}).}

(2) $\forall \, u\in C_0^{\infty}({\mathbb R}^2)$ we have
\[
\Re\langle u, H_2(\epsilon)u\rangle = \langle u, p^2u\rangle + \omega_1^2\langle u, x_1^2u\rangle + \omega_2^2\langle u, x_2^2u\rangle\geq\langle u, p^2u\rangle.
\]
Thus, (\ref{gamma}) is satisf\/ied with $\gamma=0$, $a=1$, $b=0$.

(3) Let $z\in{\mathbb C}$ and $\omega:={\rm min}\{\omega_1, \omega_2\}$. Let $u\in D(H_2(\epsilon))$ be such that $\|u\|=1$ and $u(x)=0$ for $|x|<n$, $n\in{\mathbb N}$. Then
\begin{gather*}
|z - \langle u, H_2(\epsilon)u\rangle|\geq |\langle u, H_2(\epsilon)u\rangle| - |z|\geq \Re \langle u, H_2(\epsilon)u\rangle -|z|
\nonumber\\
\phantom{|z - \langle u, H_2(\epsilon)u\rangle|}{} \geq \omega_1^2\langle u, x_1^2u\rangle + \omega_2^2\langle u, x_2^2u\rangle - |z| \geq  \omega^2\langle u, |x|^2u\rangle - |z| \geq \omega^2n^2 - |z| .\nonumber
\end{gather*}
Thus, ${\rm dist}(z, {\cal N}_n(\epsilon))\geq\omega^2n-|z|$, $\forall\, \epsilon\in{\mathbb R}$, $\forall\, n\in{\mathbb N}$, whence $\lim\limits_{n\to {\infty}}d_n(z, H(\epsilon))=+\infty$, $\forall\, \epsilon\in{\mathbb R}$ and this proves~(\ref{distance}).
\end{proof}

\begin{corollary}
\label{cor1}
Near each unperturbed eigenvalue $\lambda_{n_1, n_2}(0)$, $n_1, n_2\in{\mathbb N}_0$, of $H_2(0)$ there exists one and only one eigenvalue $\lambda_{n_1, n_2}(\epsilon)$ of $H_2(\epsilon)$ for $\epsilon\in{\mathbb R}$, $|\epsilon|$ small, converging to $\lambda_{n_1, n_2}(0)$ as $\epsilon\to 0$. Moreover $\lambda_{n_1, n_2}(\epsilon)$ is real.
\end{corollary}

\begin{proof} The f\/irst statement follows from Proposition \ref{criterion} (applicable after Proposition~\ref{prop2}), the def\/inition of stability of eigenvalues and the fact that each eigenvalue $\lambda_{n_1, n_2}(0)$ of $H_2(0)$ is simple. The reality of $\lambda_{n_1, n_2}(\epsilon)$ follows from its uniqueness, as anticipated in Remark~\ref{R1}.
\end{proof}

\section[The operator family $H_3(\epsilon)$: stability and reality of the perturbed eigenvalues]{The operator family $\boldsymbol{H_3(\epsilon)}$: stability and reality\\ of the perturbed eigenvalues}
\label{section3}

We f\/irst consider the minimal operator in $L^2({\mathbb R})$ def\/ined on $C_0^{\infty}({\mathbb R})$ by the formal expression~(\ref{Tateo}) for $|\epsilon|<1$, i.e.\ $\forall\, u\in C_0^{\infty}({\mathbb R})$ we set
\begin{equation}
\label{Tateo2}
H_3(\epsilon)u = -u'' + x^2(ix)^{\epsilon}u.
\end{equation}
Then (\ref{Tateo2}) can be rewritten as follows
\begin{gather*}
H_3(\epsilon)u = -u'' + e^{({\rm sign}\,x)i\epsilon\frac{\pi}{2}}|x|^{2+\epsilon}u
= -u'' + \cos\left(\tfrac{\pi}{2}\epsilon\right)|x|^{2+\epsilon}u + ({\rm sign}\,x)i\sin\left(\tfrac{\pi}{2}\epsilon\right)|x|^{2+\epsilon}u,
\end{gather*}
where
${\rm sign}\, x=\left\{\begin{array}{cc} 1, & {\rm if}\  x\geq 0
\\
-1, & {\rm if}\   x< 0.
\end{array}\right.$

Then $H_3(\epsilon)$ is closable and the domain of its closure (still denoted $H_3(\epsilon)$) is $D(H_3(\epsilon))=H^2({\mathbb R})\cap D(|x|^{2+\epsilon})$, $\forall\, \epsilon$: $|\epsilon|<1$. Moreover the closed operator $H_3(\epsilon)$ has compact resolvent and therefore discrete spectrum, and it is ${\cal P}{\cal T}$ symmetric if ${\cal T}$ is, once again, the complex conjugation operator and ${\cal P}$ is the parity operator def\/ined by $({\cal P} u)(x)=u(-x)$, $\forall\, u\in L^2({\mathbb R})$.

The unperturbed operator $H_3(0) = p^2 + x^2$, where $p^2=-\frac{d^2}{dx^2}$, corresponds to the one-dimensional harmonic oscillator and its eigenvalues $\lambda_n(0) = 2n+1,\, n\in{\mathbb N}_0$, are simple. Let us now proceed in analogy with Section~\ref{section2} and prove that the operator family  $H_3(\epsilon)$ satisf\/ies conditions (1)--(3).

\begin{proposition}
\label{prop3}
Conditions $(1)$--$(3)$ are satisfied by the operator family $H_3(\epsilon)$, $-1<\epsilon<1$.
\end{proposition}

\begin{proof}
(1) $\forall\, u\in C_0^{\infty}({\mathbb R})$ let $K={\rm supp}(u)$. Then
\begin{equation}\label{cont3}
\|H_3(\epsilon)u - H_3(0)u\|^2 = \int_K\big(e^{({\rm sign}\, x)i\epsilon\frac{\pi}{2}}|x|^{2+\epsilon} - x^2\big)^2|u(x)|^2 dx .
\end{equation}
Since $K$ is compact there exists a constant $c>0$ such that the integrand expression in the right hand side of (\ref{cont3}) can be bounded from above by $c|u(x)|^2$, $\forall\, x\in K$, $\forall\,\epsilon\in]{-}1,1[$. Then, by Lebesgue's dominated convergence theorem the r.h.s.\ of (\ref{cont3}) converges to zero as $\epsilon\to0$ and this proves~(\ref{continuityA}). A similar argument proves~(\ref{continuityB}).

(2) $\forall\, u\in C_0^{\infty}({\mathbb R}^2)$ we have
\[
\Re\langle u, H_3(\epsilon)u\rangle = \langle u, p^2u\rangle +  \cos\left(\tfrac{\pi}{2}\epsilon\right)\langle u, |x|^{2+\epsilon}u\rangle \geq\langle u, p^2u\rangle.
\]
Thus, as for the case of $H_2(\epsilon)$, (\ref{gamma}) is satisf\/ied with $\gamma, b=0$ and $a=1$.

(3) Again, in analogy with the argument used for $H_2(\epsilon)$, let $z\in{\mathbb C}$ and $u\in D(H_3(\epsilon))$ be such that $\|u\|=1$ and $u(x)=0$ for $|x|<n$, $n\in{\mathbb N}$. Then
\[
|z - \langle u, H_3(\epsilon)u\rangle|\geq \Re \langle u, H_3(\epsilon)u\rangle -|z|
\geq\cos\left(\tfrac{\pi}{2}\epsilon\right)\langle u, |x|^{2+\epsilon}u\rangle - |z| \geq  \tfrac{n^2}{2} - |z|
\]
for $|\epsilon|<\epsilon_1:=\frac{2}{3}$.
\end{proof}
Now, with an argument analogous to that used to prove Corollary~\ref{cor1} we obtain the following result.

\begin{corollary}
\label{cor2}
Near each eigenvalue $\lambda_n(0)=2n+1$, $n\in{\mathbb N}_0$, of $H_3(0)$ there exists one and only one eigenvalue $\lambda_n(\epsilon)$ of $H_3(\epsilon)$ for $\epsilon\in]{-}1,1[$, $|\epsilon|$ small, converging to $\lambda_n(0)$ as $\epsilon\to 0$. Moreover $\lambda_n(\epsilon)$ is real.
\end{corollary}

\section{Conclusions}
\label{section4}

Concerning Section~\ref{section1} we remark that $2\times 2$ matrices can characterize complexif\/ied classical mechanical systems (see e.g.~\cite{BeHo}) related to Schr\"odinger operators (\ref{oscillatori}) for $r=s=1$, i.e.
\[
p_1^2 + p_2^2 + \omega_1^2x_1^2 + \omega_2^2x_2^2 + i\epsilon x_1x_2.
\]
Indeed the classical equation of motion can be written as
\begin{equation}
\label{motion}
\left(\begin{array}{c} \ddot {x}_1
\\
\ddot{x}_2
\end{array}\right) = - 2\left(\begin{array}{cc} 2\omega_1^2 & i\epsilon
\\
i\epsilon & 2\omega_2^2
\end{array}\right) \left(\begin{array}{c} x_1
\\
x_2
\end{array}\right),
\end{equation}
where $\ddot{x}_k$, $k=1, 2$, denotes the second time derivative of $x_k$. It is straightforward to study the eigenvalue problem for the $2\times 2$ matrix in~(\ref{motion}), which corresponds to~(\ref{matrix}) with $\epsilon_k=2\omega_k^2$, $k=1, 2$. The eigenvalues are
\[
\lambda_{\pm}(\epsilon) = \big(\omega_1^2+\omega_2^2\big) \pm \sqrt {\big(\omega_1^2-\omega_2^2\big)^2 - \epsilon^2}.
\]
We see that the condition $|\epsilon|<|\omega_1^2-\omega_2^2|$, which yields the reality of the spectrum in quantum mechanics, also yields the reality of the frequencies of the normal oscillation modes in classical mechanics. The quantum eigenvalues are given by
\[
\lambda_{n_1, n_2}(\epsilon) = (2n_1+1)\lambda_+(\epsilon)+(2n_2+1)\lambda_-(\epsilon) ,\qquad\forall\, n_1, n_2\in{\mathbb N}_0.
\]
Expanding the functions $\lambda_{\pm}(\epsilon)$ in powers of $\epsilon$ we obtain the Rayleigh--Schr\"odinger perturbation expansion (RSPE) for the eigenvalues $\lambda_{n_1, n_2}(\epsilon)$. All expansions clearly converge for \mbox{$|\epsilon|<|\omega_1^2-\omega_2^2|$}, therefore the radius of convergence of the RSPE coincides with the threshold of transition between real and complex spectrum.

In a similar way one proves that the complexif\/ied classical Hamiltonian corresponding to the $2\times 2$ matrix (\ref{matrix2}) for $e, b\in{\mathbb R}${\samepage
\[
p_1^2 + p_2^2 +\tfrac{1}{2}(e+i\epsilon)x_1^2 + \tfrac{1}{2}(e-i\epsilon)x_2^2 + b x_1x_2
\]
admits real normal modes with real frequencies if $|\epsilon|<|b|$.}

As far as Section~\ref{section2} is concerned we remark that a suitable generalization regards polynomial perturbations of the $d$-dimensional harmonic oscillator, $d>2$. In addition, still in dimension $d=2$ one could try to study the case of resonant frequencies, not with stability methods (which fail because of the degeneracy of the unperturbed eigenvalues), but proving the reality of the Rayleigh--Schr\"odinger perturbation expansion and its summability.

Finally, concerning Section~\ref{section3} an open question is the rigorous proof of the existence of complex eigenvalues for $-1<\epsilon<0$, supporting the existing numerical results, and the analysis of the case $\epsilon<-1$.

\subsection*{Acknowledgement}

We wish to thank R.~Tateo for useful correspondence.

\pdfbookmark[1]{References}{ref}
\LastPageEnding

\end{document}